\documentclass{llncs}

\usepackage{amsmath}
\usepackage{amssymb}
\usepackage{algorithm}
\usepackage{bbding}
\usepackage{booktabs}
\usepackage{cite}
\usepackage{hyperref}
\usepackage{cleveref}
\usepackage{fancyvrb}
\usepackage[T1]{fontenc}
\usepackage{graphicx}
\usepackage{algpseudocode}
\usepackage{mathtools}
\usepackage{tikz}
\usepackage{upgreek}
\usepackage{verbatim}
\usepackage{xcolor}

\usetikzlibrary{positioning, arrows.meta, shapes.geometric}
\graphicspath{{./include/}}
\newcommand{\heurOriginal}{\ensuremath{\mathrm{Original}}}
\newcommand{\heurRandom}{\ensuremath{\mathrm{Random}}}
\newcommand{\heurSU}{\ensuremath{\mathrm{SU}}}
\newcommand{\heurCumulSU}{\ensuremath{\mathrm{CumulSU}}}
\newcommand{\heurCumulPred}{\ensuremath{\mathrm{CumulPred}}}
\newcommand{\heurNumSucc}{\ensuremath{\mathrm{NumSucc}}}
\newcommand{\heurLastSucc}{\ensuremath{\mathrm{LastSucc}}}
\newcommand{\pred}{\mathrm{pred}}
\newcommand{\gsucc}{\mathrm{succ}}

\title{Optimising Metamath Proofs for Human Working Memory}
\author{
  Jeremy Lindsay\textsuperscript{\Envelope}\orcidID{0009-0003-9584-1207} \and
  Cezary Kaliszyk \orcidID{0000-0002-8273-6059} \and
  Christine Rizkallah \orcidID{0000-0003-4785-2836}
}
\institute{
  University of Melbourne, Australia \\
  \email{
    jeremy.n.lindsay@student.unimelb.edu.au \\
    cezary.kaliszyk@unimelb.edu.au \\
    christine.rizkallah@unimelb.edu.au
}}

\begin{document}

\maketitle

\begin{abstract}
  Mathematical proofs vary in legibility. While most proof optimisation techniques seek to minimise proof size, the strategic reordering of inferences can reduce the working memory demand of proof checking without altering overall size. Metamath serves as a prime case study for this approach: its verification architecture requires proof steps to be ordered in a manner that prioritises algorithmic efficiency over readability. In this paper, we introduce algorithms to minimise both peak and cumulative memory consumption, applying the latter as a novel proxy for sustained human cognitive effort. We achieve this by representing proofs as directed acyclic graphs and modelling their execution as a pebbling game. Finding an optimal ordering via brute force is computationally infeasible, so we use heuristics to provide approximations. We apply these algorithms across Metamath's ZFC set theory library and present case studies demonstrating how automated reordering systematically improves the presentation of formal mathematics.

\end{abstract}

\section{Introduction}

An important aspect of proof readability is the demand it places on a
reader's working memory.
Consider a mathematician
presenting a proof on a chalkboard, writing down intermediate
results and erasing them when they are no longer needed.
The total amount of
chalk used reflects the proof's length,
and the size of the board limits the number of results that can
be simultaneously retained for later reference.
Yet, if even a small chalkboard is continuously cluttered with
results awaiting later reference,
the audience may become overwhelmed.
We, therefore, prefer proofs
that are not only short and space-efficient, but that also
keep the running total of
retained intermediate results low.
In this
paper, we introduce algorithms to reorder proof steps that minimise
the latter two metrics, while preserving the proof length.
Though legibility is subjective and multifaceted,
we target these metrics as concrete, quantifiable proxies for human
cognitive load.

Formal mathematics libraries of proof assistants like
Lean~\cite{lean_2021}, Rocq~\cite{rocq_2025}, and Metamath~\cite{metamath}, provide a natural setting for this
problem. Their proofs have
explicit dependency structure, so valid reorderings can be specified
precisely rather than judged informally.
The methods we present in this paper are general-purpose
and require only abstract graph-based representations of proof
structure. Nevertheless, we perform our experiments on Metamath proofs
in particular, for a few reasons. First, Metamath's syntax and verification
mechanisms are simple, making it straightforward to export proofs into a
graph format.
Recent tooling such as LeanTree \cite{leantree_2025} demonstrates the
feasibility of extracting structured proof representations from
other proof assistants, but we leave this for future work.
Second, Metamath requires proof steps to be ordered in a
highly specific manner that is optimised for verification rather than
readability. By default, Metamath proof editors display proofs in this
native format, though they allow steps to be reordered during the
editing stage. Thus Metamath provides a clear testbed for studying whether
automatic reordering can improve the presentation of formal proofs.

Our approach is as follows. We represent a proof as a
\emph{directed acyclic graph (DAG)} whose
vertices are distinct inferences
and edges represent logical dependencies.
A traditional,
sequential presentation of the proof is then a
\emph{topological order} on this DAG. To model memory demand
as the proof is followed step by step, we use the formalism of \emph{pebbling games}. It has been shown for a variety of metrics that brute force
methods for computing an optimal topological order are infeasible in
practice~\cite{pak_np_hard_2015,fellner_greedy_2019,sethi_complete_1973}. We therefore
turn to heuristic algorithms.

Our contributions are the following:
\begin{itemize}
\item We formulate the problem of optimising proofs for
  \emph{sustained} demand on working memory. This is captured by
   cumulative memory use across the proof, which we measure alongside
   the better-studied objective of minimising peak memory consumption.
\item We introduce metrics closely related to Karol P\k{a}k's
  legibility metrics~\cite{pak_np_hard_2015,pak_algorithms_2010},
  and show where our memory consumption metrics diverge.
\item We introduce a novel variant of the pebbling game that treats
  unpebbling as an automatic process, simplifying memory calculations
  as a result.
\item We develop heuristic reordering methods that combine ideas from
  Fellner and Paleo's greedy algorithm~\cite{fellner_greedy_2019},
  Sethi--Ullman numbering~\cite{sethi_generation_1970}, and other
  lightweight ordering criteria. We apply
  these algorithms to Metamath's ZFC set theory library
  \texttt{set.mm}, analyse the results, and present
  case studies.
\end{itemize}

More broadly, this contributes towards improving proof legibility for humans,
a problem that is likely to become more important as LLM-based proof
generation and proof transformation become more common~\cite{kimina_prover_2025,deepseek_prover_2_2025,gallardo_formalization_2025}.

\section{Related Work}

Most proof optimisation work seeks to minimise proof length or verification
time through techniques like definition invention~\cite{vyskocil_invention_2010},
redundancy pruning~\cite{pak_algorithms_2010}, LLM-based
refactoring~\cite{ahuja_improver_2025,gu_proofoptimizer_2026,zhou_refactor_2024},
and abstraction discovery~\cite{kaliszyk_millions_lemmas_2015,rahul_proof_nodate}.
Instead, we consider the problem of how the \emph{ordering} of proof steps
affects memory consumption and legibility, while the overall proof length is
held constant.

Karol P\k{a}k was the first to consider the impact of proof step ordering on
legibility. His work encompasses novel optimisation
algorithms~\cite{pak_mizar_legibility_2014}, experimental studies with human
subjects~\cite{pak_experimental_readability_2016}, and theoretical results
showing that various proof reordering problems are
NP-complete~\cite{pak_np_hard_2015}. P\k{a}k's metrics are based on the
\emph{close reference principle}, which recommends that premises should
be introduced soon before they are required in a logical inference. These metrics
can be calculated on a static proof graph; our work diverges in that we model
the evolution of memory consumption over time during proof checking. Nevertheless,
we introduce two metrics, the \emph{mean index distance}, and the
\emph{adjacency rate}, that are inspired by P\k{a}k's
\emph{3rd Method of Improving Legibility}~\cite{pak_np_hard_2015} and the
\emph{4th Method of Improving Legibility}~\cite{pak_mizar_legibility_2014}
respectively. Using these metrics, we present case studies demonstrating how
our memory consumption metrics can diverge from
metrics based solely on the close reference principle.

Pebbling games are used in the field of \emph{proof complexity} to analyse the
space complexity of proof verification~\cite{nordstrom_pebble_2013}.
Proof complexity is a theoretical discipline concerned with establishing
asymptotic bounds on the size and space requirements of various proof
systems~\cite{krajieck_complexity_2019}. In contrast, our work is applied:
rather than analysing the theoretical limits of Metamath, our focus is the
efficient optimisation of concrete proofs. The work of Fellner and Paleo
\cite{fellner_greedy_2019} most closely resembles ours: they study
space optimisation for machine-generated resolution proofs using both an exact
SAT formulation and a greedy algorithm. Our depth-first search approach is
directly inspired by the latter, but they focus only on peak memory
consumption, whereas we view cumulative memory consumption as a proxy for
sustained cognitive load in humans.

\section{Metamath}

Metamath is a logic-agnostic proof assistant: without provided axioms, it contains
no inherent mathematical content.
Instead, users specify Hilbert-style axioms as explicit foundations for different mathematics
libraries. In this paper, we restrict ourselves to the ZFC set theory
library \texttt{set.mm}, the largest and most mature Metamath library.

During verification, Metamath uses a \emph{stack} to keep track of
proof states.
Each proof step either introduces a given assumption or
applies an inference rule, producing a new mathematical
expression.\footnote{This is a simplification: like the Metamath Proof Explorer, we abstract away purely syntactic construction steps and track only expressions with typecode \texttt{|-}.} In
\texttt{set.mm}, \emph{modus ponens} appears as the axiom \texttt{ax-mp},
which states that given $P$ and $P \to Q$, we can derive
$Q$. Crucially, Metamath theorems, like functions in programming
languages, expect arguments to be specified in a particular order.
Theorems are assigned abbreviated mnemonic labels:
\texttt{mp2}, for instance, stands for \emph{double modus ponens}, and it has three
ordered premises:
\begin{center}
  \setlength{\tabcolsep}{10pt}
  \begin{tabular}{l|l|l}
    premise 1
    & \texttt{mp2.1}
    & $\varphi$ \\
    premise 2
    & \texttt{mp2.2}
    & $\psi$ \\
    premise 3
    &\texttt{mp2.3}
    & $\varphi \to (\psi \to \chi)$ \\
    conclusion
    & \texttt{mp2}
    & $\chi$ \\
  \end{tabular}
\end{center}

To prove \texttt{mp2}, we apply modus ponens twice. Metamath's stack, as depicted in \Cref{tab:mp2-stack}, evolves as follows:
  \begin{enumerate}
    \item Initially, the stack is empty.
    \item Premise 2, $\psi$, is pushed to the stack.
    \item Premise 1, $\varphi$, is pushed.
    \item Premise 3, $\varphi \to (\psi \to \chi)$, is pushed.
    \item Modus ponens is applied to $\varphi$ and $\varphi \to (\psi \to \chi)$, popping both and pushing the result $\psi \to \chi$.
    \item Modus ponens is applied again to $\psi$ and $\psi \to \chi$, popping both and pushing the result $\chi$.
      The stack now contains exactly the goal formula, so the proof is complete.
  \end{enumerate}
In general, the application of a theorem with $n$ premises
corresponds to mechanically popping the top $n$ elements of the
stack, applying the theorem, and pushing the result onto the stack.

  \begin{table}[htbp]
    \caption{Metamath's stack during the proof of \texttt{mp2}. Each column corresponds to a step in the proof, and each row corresponds to a position on the stack.}
    \centering
    \renewcommand{\arraystretch}{1.3}
    \setlength{\tabcolsep}{7pt}
    \begin{tabular}{|c|c|c|c|c|c|}
      \hline
      & & & $\varphi \to (\psi \to \chi)$ & & \\
      \hline
      & & $\varphi$ & $\varphi$ & $\psi \to \chi$ & \\
      \hline
      & $\psi$ & $\psi$ & $\psi$ & $\psi$ & $\chi$ \\
      \hline \hline
      \textbf{1} & \textbf{2} & \textbf{3} & \textbf{4} & \textbf{5} & \textbf{6} \\
      \hline
    \end{tabular}
    \label{tab:mp2-stack}
  \end{table}

Metamath requires inferences to be ordered in the precise sequence in which they are
evaluated on the stack. (If an expression is required multiple times in a proof,
Metamath will cache it and push it back onto the stack when necessary.
We do not consider this operation a new inference.)
This makes the native Metamath format unusually laborious to read.
The proof of \texttt{prmunb}
(unboundedness of primes)
is paradigmatic:
the first step introduces an expression that is not referenced until
the very end of the proof, a full 47 steps later.
In practice, proof editors like \texttt{mmj2} can be used to present proof steps in any order,
so long as their premises are referenced appropriately.
However, proofs are initially loaded from \texttt{set.mm} with the
original step order, and are also compiled back into the native format.

\section{Proof DAGs and Topological Orders}

In this section, we show how proofs can be represented as directed acyclic
graphs (DAGs), with vertices representing inferences,
and edges representing logical dependencies.
The acyclic condition prohibits circular reasoning,
whereby a mathematical expression could be derived from itself.

\begin{definition}[Proof DAG]
  \label{def:proof-dag}
  Let $\mathcal{E}$ denote the set of all valid mathematical expressions, and let $P$ be a formal proof consisting of a sequence of inferences.
  A \emph{proof DAG} is a tuple $(V, E, \varepsilon, \lambda)$ where:
  \begin{enumerate}
    \item The pair $(V, E)$ is a directed acyclic graph with a unique sink. The vertices $V$ represent the distinct inferences in $P$, and an edge $(u, v) \in E$ indicates that inference $v$ directly consumes the output of inference $u$.
    \item The function $\varepsilon : V \to \mathcal{E}$ maps each vertex to the mathematical expression it derives. (Injectivity is not required, so expressions may be rederived.)
    \item The function $\lambda : E \to \mathbb{Z}^+$ assigns to each edge $(u, v) \in E$ the premise index of $u$ within the theorem applied at $v$.
  \end{enumerate}
\end{definition}

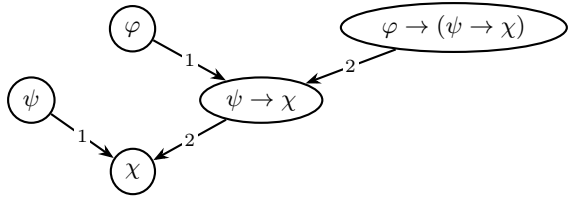
\begin{figure}[htbp]
  \centering
  \begin{tikzpicture}[
      font=\footnotesize,
      node distance=5mm and 9mm,
      vertex/.style={
        draw,
        thick,
        ellipse,
        align=center,
        inner xsep=3pt,
        inner ysep=2pt,
        minimum height=6mm
      },
      edge/.style={->, >={Stealth[length=2mm]}, thick}
    ]

    \node[vertex] (n11) {$\chi$};
    \node[vertex, above left=of n11] (n3) {$\psi$};
    \node[vertex, above right=of n11] (n10) {$\psi \to \chi$};
    \node[vertex, above left=of n10] (n8) {$\varphi$};
    \node[vertex, above right=of n10] (n9) {$\varphi \to (\psi \to \chi)$};

    \draw[edge] (n8) -- node[midway, fill=white, inner sep=1pt] {\scriptsize 1} (n10);
    \draw[edge] (n9) -- node[midway, fill=white, inner sep=1pt] {\scriptsize 2} (n10);
    \draw[edge] (n3) -- node[midway, fill=white, inner sep=1pt] {\scriptsize 1} (n11);
    \draw[edge] (n10) -- node[midway, fill=white, inner sep=1pt] {\scriptsize 2} (n11);

  \end{tikzpicture}
  \caption{Proof DAG for the double modus ponens theorem (\texttt{mp2}). Vertices $v$ represent distinct inferences and are labelled with their derived expressions $\varepsilon(v)$.}
  \label{fig:mp2-dag}
\end{figure}

In \Cref{fig:mp2-dag}, for example, the three sources are mapped
by $\varepsilon$ to the three premises of \texttt{mp2},
and the sink is mapped to the conclusion $\chi$.
The edges are labelled by $\lambda$ according to \texttt{ax-mp}'s
ordered premises: the minor premise $P$ followed by
the major premise $P \to Q$.

Given a proof DAG, there are in general multiple valid sequences
(linearisations) of proof steps that can be constructed.
Each of these corresponds to a \emph{topological order} $\prec$ on the DAG;
we write $V_\prec$ for the induced sequence of vertices in $V$.
Since an edge $(u, v) \in E$ represents a logical dependency
on $u$ by $v$, ideally $u$ should be executed not long before $v$.
The average of this dependency distance across all edges is thus
a natural proxy metric for legibility.

\begin{definition}[Mean index distance]
  Let $G = (V, E)$ be a DAG, and let $\prec$ be a topological order on $G$. The \emph{index distance} of any two vertices $u, v \in V$ is defined as $d_\prec(u, v) \coloneqq \left\lvert \mathrm{idx}_\prec(u) - \mathrm{idx}_\prec(v) \right\rvert$, where $\mathrm{idx}_\prec(v)$ is the index of $v$ in $V_\prec$. The \emph{mean index distance (MID)} of $G$ with respect to $\prec$ is
    $$
      \mathrm{MID}_\prec(G) \coloneqq
      \frac{1}{\lvert E \rvert}
      \sum_{(u, v) \in E} d_\prec(u, v) .
    $$
\end{definition}

In the limit, legibility is optimised when $d_\prec(u, v) = 1$
for every edge $(u, v) \in E$. The following definition captures the extent
to which this ideal is achieved.

\begin{definition}[Adjacency rate]
  Let $G = (V, E)$ be a DAG, and let $\prec$ be a topological order on $G$. The \emph{adjacency rate} of $G$ with respect to $\prec$ is defined as
  $$
    \mathrm{AdjRate}_\prec(G) \coloneqq
    \frac
    {\lvert \{ (u, v) \in E : d_\prec(u, v) = 1 \} \rvert}
    { \lvert E \rvert} .
  $$
\end{definition}

\section{Pebbling Games}

To compare the working memory used while checking different linearisations
of the same proof, we need a way to track which intermediate expressions
must remain available at
each step.
The formalism of \emph{pebbling games} played on a DAG provides a
natural abstraction for this.
Intuitively, placing a pebble on a vertex $v$ indicates that the expression $\varepsilon(v)$ that it derives is currently stored in working memory. Only after this expression is no longer required should $v$ be ``unpebbled'' and released from memory.
From this formalism, we can derive metrics that model
\emph{peak} and \emph{cumulative} memory
demand across the whole proof.

  \begin{definition}[Pebbling game, adapted from Nordström \cite{nordstrom_pebble_2013}]
    \label{def:pebbling-game}
    Let $G = (V, E)$ be a DAG. The \emph{pebbling game} on $G$ is a one-player game where each legal move is a pebbling or unpebbling of exactly one vertex as follows:
    \begin{enumerate}
      \item Any unpebbled source may be pebbled.
      \item Any unpebbled vertex whose predecessors are pebbled may be pebbled.
      \item Any pebbled vertex may be unpebbled.
    \end{enumerate}
    To begin, the DAG is entirely unpebbled, and the goal of the game is to pebble all sinks. Formally, a \emph{pebbling strategy} for $G$ is a sequence $P = (P_0, P_1, \ldots, P_n)$ of \emph{pebble configurations} $P_t \in \mathcal P(V)$ such that $P_0 = \emptyset$ and $P_n = S$, where $S \subseteq V$ is the set of all sinks; and such that each $P_t$ follows from $P_{t - 1}$ according to a legal move. The \emph{cost} of a configuration $P_t$ is $\lvert P_t \rvert$, the number of active pebbles on the DAG.
\end{definition}

When $G$ is a proof DAG, the goal is to pebble the unique sink, i.e.,
its conclusion.
Consider \Cref{fig:mp2-dag}. Before each application of modus ponens,
its two premises must be pebbled; the game ends once $\chi$ is
pebbled.
Standard pebbling, however, leaves
the timing of unpebbling to the player. For proof linearisation this
is an unnecessary degree of freedom: once an order is fixed, we want
memory usage to be determined by which expressions still have future
uses. This motivates the following restricted variant.

\begin{definition}[One-shot pebbling game with implicit unpebbling] \label{def:pebbling-game-variant}
    \Cref{def:pebbling-game} is augmented with the following rules to form a variant of the pebbling game.
    \begin{enumerate}
      \item (\emph{One-shot}.) Each vertex can be pebbled at most once.
      \item (\emph{Implicit unpebbling}.) If, after a move, all successors of a vertex $v$ have at some point been pebbled, then $v$ is automatically unpebbled. This does not count as a move.
    \end{enumerate}
\end{definition}

The one-shot rule stipulates that each inference should be
evaluated at most once in a proof.
We view the strategic re-evaluation of past inferences as a separate
theoretical problem and prohibit it here.
(If a proof inefficiently derives the same mathematical expression via
two distinct inferences, e.g., $\varepsilon(v_1) = \varepsilon(v_2)$,
our model still permits both to be pebbled.)
The implicit unpebbling rule captures the idea that premises do not (in
general) need to be held in memory together with the conclusion once
it is established. By analogy, this is like erasing premises
on the chalkboard to make space for the conclusion.

There are caveats to using pebbling games for measuring cognitive load.
First, in the chalkboard analogy,
they measure demand on the \emph{audience's} memory rather than the
presenter's---the presenter has decided in advance how long each mathematical expression must
stay on the board, relieving the audience of having to perform explicit memory management.
In other words, pebbling games use ``live mathematical expressions''
only as a proxy for working memory load.
Second, in cases where the optimal space or cumulative cost of a proof is very high,
it is debatable whether readability can be noticeably improved at all.
Nevertheless, we speculate that optimising proofs with respect to these metrics
improves dataset quality for neural and LLM-based methods.

Given a pebbling strategy, two metrics are of particular interest: the
\textit{maximum}, and the \textit{cumulative} cost across all
configurations.
  \begin{definition}[Pebbling space and cumulative pebbling cost]
    Let $G = (V, E)$ be a DAG, and let $P$ be a pebbling strategy on $G$. The \emph{pebbling space} of $P$, and the \emph{cumulative pebbling cost} of $P$, are defined respectively as
    $$
      \mathrm{sp}(P) \coloneqq \max_t \lvert P_t \rvert
      \qquad \text{and} \qquad
      \mathrm{cc}(P) \coloneqq \sum_t \lvert P_t \rvert .
    $$
\end{definition}

\begin{figure}[htbp]
    \centering
    \begin{minipage}[t]{0.48\textwidth}
      \centering
      \begin{tikzpicture}[
          font=\footnotesize,
          node distance=3mm and 3mm,
          vertex/.style={
            draw,
            line width=0.6pt,
            circle,
            align=center,
            inner sep=1pt,
            minimum size=5mm
          },
          edge/.style={->, >={Stealth[length=2mm]}, thick}
        ]

        \node[vertex] (e) {$E$};
        \node[vertex, above left=of e] (a) {$A$};
        \node[vertex, above right=of e] (d) {$D$};
        \node[vertex, above left=of d] (b) {$B$};
        \node[vertex, above right=of d] (c) {$C$};

        \draw[edge] (b) -- (d);
        \draw[edge] (c) -- (d);
        \draw[edge] (a) -- (e);
        \draw[edge] (d) -- (e);
      \end{tikzpicture} \\
      {\small(a)}
    \end{minipage}\hfill
    \begin{minipage}[t]{0.48\textwidth}
      \centering
      \begin{tikzpicture}[
          font=\footnotesize,
          node distance=3mm and 3mm,
          vertex/.style={
            draw,
            line width=0.6pt,
            circle,
            align=center,
            inner sep=1pt,
            minimum size=5mm
          },
          edge/.style={->, >={Stealth[length=2mm]}, thick}
        ]
        \node[vertex] (d2) {$D$};
        \node[vertex, above left=of d2] (a2) {$A$};
        \node[vertex, above right=of d2] (c2) {$C$};
        \node[vertex, above right=of c2] (b2) {$B$};
        \draw[edge] (a2) -- (d2);
        \draw[edge] (b2) -- (c2);
        \draw[edge] (c2) -- (d2);
      \end{tikzpicture} \\
      {\small(b)}
    \end{minipage}
    \caption{Two simple DAGs. DAG (a) has the same structure as that of \texttt{mp2} in \Cref{fig:mp2-dag}.}
    \label{fig:two-simple-dags}
  \end{figure}
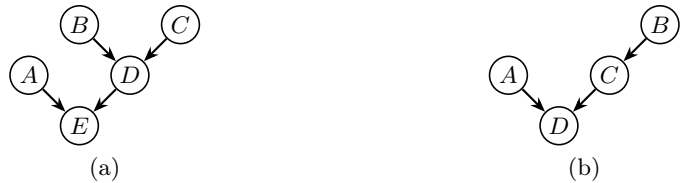

\Cref{tab:dag-space-vs-space,tab:space-vs-cost} demonstrate
calculations of space and cumulative cost for the two DAGs in
\Cref{fig:two-simple-dags}. Two pebbling strategies with different
pebbling spaces are presented for \Cref{fig:two-simple-dags}~(a). For \Cref{fig:two-simple-dags}~(b), two strategies are presented that require
the same space but have different cumulative costs.
Thus optimising for space and optimising for
cumulative cost can lead to different strategies.

  \begin{table}[htbp]
    \caption{Two possible pebbling strategies for \Cref{fig:two-simple-dags}~(a). Strategy 1 requires space 3 whereas strategy 2 requires space 2.}
    \centering
    \renewcommand{\arraystretch}{1.2}
    \setlength{\tabcolsep}{7pt}
    \begin{tabular}{|c||c|l|c||c|l|c|}
      \hline
      \textbf{Step}
      & $\prec_1$
      & \textbf{Configuration}
      & \textbf{Cost}
      & $\prec_2$
      & \textbf{Configuration}
      & \textbf{Cost} \\
      \hline
      1 & A & $\{ A \}$ & 1 & B & $\{ B \}$ & 1 \\ \hline
      2 & B & $\{ A, B \}$ & 2 & C & $\{ B, C \}$ & 2 \\ \hline
      3 & C & $\{ A, B, C \}$ & 3 & D & $\{ D \}$ & 1 \\ \hline
      4 & D & $\{ A, D \}$ & 2 & A & $\{ D, A \}$ & 2 \\ \hline
      5 & E & $\{ E \}$ & 1 & E & $\{ E \}$ & 1 \\ \hline
    \end{tabular}
    \label{tab:dag-space-vs-space}
  \end{table}

  \begin{table}[htbp]
    \caption{Two possible pebbling strategies for \Cref{fig:two-simple-dags}~(b). Both require space 2, but strategy 1 has cumulative cost 6 while strategy 2 has cumulative cost 5.}
    \centering
    \renewcommand{\arraystretch}{1.2}
    \setlength{\tabcolsep}{7pt}
    \begin{tabular}{|c||c|l|c||c|l|c|}
      \hline
      \textbf{Step}
      & $\prec_1$
      & \textbf{Configuration}
      & \textbf{Cost}
      & $\prec_2$
      & \textbf{Configuration}
      & \textbf{Cost} \\
      \hline
      1 & A & $\{ A \}$ & 1 & B & $\{ B \}$ & 1 \\ \hline
      2 & B & $\{ A, B \}$ & 2 & C & $\{ C \}$ & 1 \\ \hline
      3 & C & $\{ A, C \}$ & 2 & A & $\{ C, A \}$ & 2 \\ \hline
      4 & D & $\{ D \}$ & 1 & D & $\{ D \}$ & 1 \\ \hline
    \end{tabular}
    \label{tab:space-vs-cost}
  \end{table}

  Our proof optimisation algorithms do not compute pebbling strategies directly, but rather topological orders. Each topological order on a DAG, however, induces the obvious pebbling strategy as follows.

  \begin{definition}[Canonical pebbling strategy]
    Let $G = (V, E)$ be a DAG, and let $\prec$ be a topological order on $G$ such that $V_\prec = (v_1, v_2, \ldots, v_n)$. The \emph{canonical pebbling strategy} $P_\prec$ is the unique pebbling strategy constructed as follows:
    \begin{itemize}
      \item $P_0 \coloneqq \emptyset$
      \item $P_t \coloneqq \mathrm{UnPeb} \left( P_{t - 1} \cup \{ v_t \} \right)$ for
        $t = 1, 2, \ldots, n$,
    \end{itemize}
    where $\mathrm{UnPeb} : \mathcal P(V) \to \mathcal P(V)$ unpebbles vertices according to the implicit unpebbling rule in \Cref{def:pebbling-game-variant}.
\end{definition}

\section{Proof Reordering Algorithms}

Given a proof DAG, our goal is to find a pebbling strategy whose space
and cumulative cost is minimised. In this section we present
\Cref{alg:bottom-up-dfs}, a bottom-up depth-first search (DFS) approach
adapted from Fellner and Paleo~\cite{fellner_greedy_2019}.
During search, the order in which the predecessors of each vertex
are visited is determined by an \emph{ordering policy}.

\begin{definition}[Ordering policy]
  Let $G = (V, E, \varepsilon, \lambda)$ be a proof DAG. An \emph{ordering policy} $\pi$ is a rule that assigns to each vertex $v \in V$ a total preorder $\geq_v^\pi$ or total order $>_v^\pi$ on $\pred(v)$. For distinct $u, w \in \pred(v)$, we define the following policies:
  \begin{center}
    \setlength{\tabcolsep}{10pt}
    \begin{tabular}{l r@{\;}c@{\;}l}
      \toprule
      \textbf{Ordering policy $\pi$ (strict)}
      & \multicolumn{3}{l}{\textbf{Condition for $u >_v^\pi w$}} \\
      \midrule
      \heurOriginal & $\lambda(u, v)$ & $<$ & $\lambda(w, v)$ \\
      \heurRandom & \multicolumn{3}{l}{Holds with probability $1 / 2$} \\
      \midrule
      \textbf{Ordering policy $\pi$ (non-strict)}
      & \multicolumn{3}{l}{\textbf{Condition for $u \geq_v^\pi w$}} \\
      \midrule
      \heurSU & $W_\heurSU(u)$ & $\geq$ & $W_\heurSU(w)$ \\
      \heurCumulSU & $W_\heurCumulSU(u)$ & $\geq$ & $W_\heurCumulSU(w)$ \\
      \heurCumulPred & $W_\heurCumulPred(u)$ & $\geq$ & $W_\heurCumulPred(w)$ \\
      \heurNumSucc & $\lvert \gsucc(u) \rvert$ & $\geq$ & $\lvert \gsucc(w) \rvert$ \\
      \heurLastSucc & $W_\heurLastSucc(u)$ & $\geq$ & $W_\heurLastSucc(w)$ \\
      \bottomrule
    \end{tabular}
  \end{center}
  where the weights $W_\pi$ are defined as follows: for all non-source vertices $v \in V$,
  \begin{align*}
    W_\heurSU(v) &\coloneqq \max_{1 \leq i \leq k} \big( W_\heurSU(u_{(i)}) + i - 1 \big) \\
    W_\heurCumulSU(v) &\coloneqq W_\heurSU(v) + \sum_{u \in \pred(v)} W_\heurCumulSU(u) \\
    W_\heurCumulPred(v) &\coloneqq 1 + \sum_{u \in \pred(v)} W_\heurCumulPred(u) \\
    W_\heurLastSucc(v) &\coloneqq \left\vert \left\{ (u, v) \in E : \lambda(u, v) > \lambda(u, w) \ \text{for all} \ w \in \gsucc(u) \right\} \right\vert.
  \end{align*}
  Here, $(u_{(1)}, \dots, u_{(k)})$ is $\pred(v)$ sorted so that $ W_\heurSU(u_{(1)}) \geq \cdots \geq W_\heurSU(u_{(k)}) $. For $\pi = \heurLastSucc$, all source weights are set to zero; for all other policies, source weights are set to one.
\end{definition}

The initialism \heurSU{} stands for \emph{Sethi-Ullman} numbering,
originally developed for efficient evaluation of binary computation
trees~\cite{sethi_generation_1970} and later generalised by Appel and
Supowit~\cite{appel1987generalizations}. The \heurLastSucc{} heuristic
is taken directly from Fellner and Paleo~\cite{fellner_greedy_2019},
which was their best-performing heuristic. We include
\heurCumulSU{} and \heurCumulPred{} as simple adaptations
meant to target cumulative cost. These ordering policies 
are driven by the intuition that it is generally better to evaluate 
heavier branches before lighter ones, thereby minimising the duration that 
intermediate results must remain in memory.
In \Cref{sec:case-studies} we present examples that support this
intuition.

Given a sequence $(\pi_1, \ldots, \pi_n)$ of ordering policies, we can
combine them lexicographically into a single policy such that later
policies break ties in earlier ones. \Cref{alg:bottom-up-dfs} requires
all input orders to be strict, which we achieve by appending the
$\heurOriginal$ policy as a final tie-breaker to all non-strict policies.

\begin{algorithm}[htbp]
  \caption{Bottom-up DFS linearisation of a DAG}
  \label{alg:bottom-up-dfs}
  \begin{algorithmic}[1]
    \Require A DAG $G = (V, E)$ with sink $s$. A family $(>_v^\pi)_{v \in V}$ of predecessor orders.
    \Ensure A list $\mathit{result}$, an ordering of $V$.
    \State $\mathit{visited} \gets \emptyset$
    \State $\mathit{result} \gets [\ ]$
    \Function{DFS}{$v$}
    \If{$v \notin \mathit{visited}$}
    \State $\mathit{visited} \gets \mathit{visited} \cup \{v\}$
    \ForAll{$u \in \pred(v)$ sorted by $>_v^\pi$ descending}
    \State \Call{DFS}{$u$}
    \EndFor
    \State append $v$ to $\mathit{result}$
    \EndIf
    \EndFunction
    \State \Call{DFS}{$s$}
    \State \Return $\mathit{result}$
  \end{algorithmic}
\end{algorithm}

\section{Experiments}

\begin{figure}
  \centering
  \includegraphics[width=0.8\textwidth]{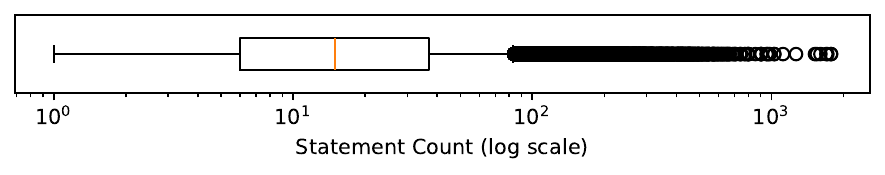}
  \includegraphics[width=0.8\textwidth]{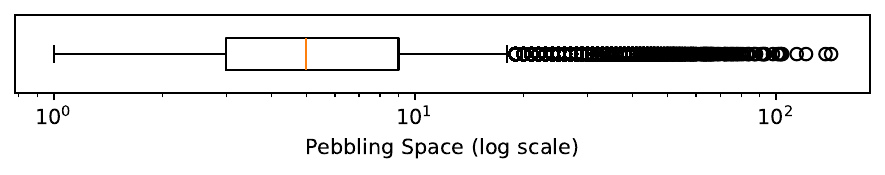}
  \includegraphics[width=0.8\textwidth]{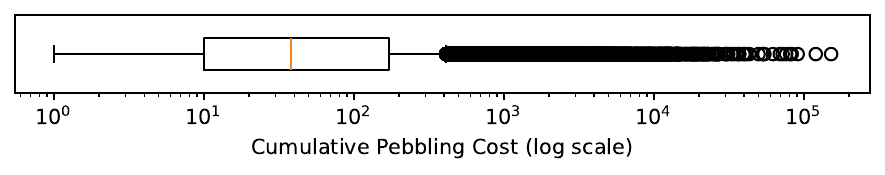}
  \caption{Box plots of statement counts, pebbling spaces, and cumulative pebbling costs of the original \texttt{set.mm} proofs (log scale). In all three cases, the distributions are heavily right-skewed: the median statement count is 15, the median space is 5, and the median cumulative cost is 38.}
  \label{fig:boxplot-metrics}
\end{figure}

Our experiments are run on Metamath's \texttt{set.mm} library~\cite{setmm_repository} at Git commit hash \texttt{d43a7bbab3145af25ea755d187f473ecff2670c2}. Rather than parsing
proofs directly, we modify the \texttt{mmverify.py} Python
verifier~\cite{wheeler_mmverify_py,lindsay_python_exporter_2026} so that at each stack operation the context is logged
to JSON.
Each proof is then converted into a DAG, on which we run \Cref{alg:bottom-up-dfs} and
calculate metrics. \Cref{fig:boxplot-metrics} presents some basic statistics
about \texttt{set.mm}.

\subsection{Methodology}

For each proof DAG $G$ in \texttt{set.mm}, we use \Cref{alg:bottom-up-dfs} to compute a
topological order $\prec$ on $G$ for each of the ordering policies below:
\begin{itemize}
  \item \heurOriginal
  \item \heurRandom\footnote{For reproducibility, the \heurRandom{} policy uses a fixed pseudorandom seed of 2026.}
  \item (\heurSU, \heurOriginal)
  \item (\heurSU, \heurCumulSU, \heurOriginal)
  \item (\heurCumulPred, \heurOriginal)
  \item (\heurNumSucc, \heurOriginal)
\end{itemize}
The \heurRandom{} ordering policy serves as a baseline for comparison: intuitively, it is
reasonable to expect it to perform similarly to Metamath's original ordering. This is
likely because the premise orderings $\lambda(e)$ of each edge $e \in E$ in a proof
DAG are effectively arbitrary.
By reordering premises in \texttt{set.mm}'s theorem definitions,
one could theoretically make \heurOriginal{} coincide with other policies
on certain proofs.

Once a topological order $\prec$ is obtained for each policy, we calculate the pebbling space $\mathrm{sp}(P_\prec)$ and cumulative pebbling cost $\mathrm{cc}(P_\prec)$ of the canonical pebbling strategy $P_\prec$. We also calculate the mean index distance $\mathrm{MID}_\prec(G)$ and adjacency rate $\mathrm{AdjRate}_\prec(G)$. For space and cumulative cost, we calculate the relative improvement as
$$
  \Updelta_\mathrm{metric} \coloneqq \frac{\mathrm{metric}(P_\heurOriginal) - \mathrm{metric}(P_\prec)}{\mathrm{metric}(P_\heurOriginal)} .
$$

\Cref{alg:bottom-up-dfs} may produce strictly worse pebbling strategies; for each ordering
policy we calculate the percentage of proofs in \texttt{set.mm} where this occurs.
At the other extreme, we evaluate the percentage of proofs where space attains the
theoretical lower bound---that is,
where $\mathrm{sp}(P_\prec) = \max_{v \in V} \lvert \pred(v) \rvert$.
Finally, we record the execution time of \Cref{alg:bottom-up-dfs} for each ordering policy
across all proofs in \texttt{set.mm}. Experiments were run with Python~3.14.3 on an
Apple M5 MacBook Pro with 16GB unified memory, running macOS~26.4 (Tahoe). The code for our experiments is available online \cite{lindsay_metamath_memory_2026}.

Sethi-Ullman numbering and its generalisation by Appel and Supowit are optimal for
trees with respect to pebbling space~\cite{sethi_generation_1970,appel1987generalizations}.
We therefore separate treelike and non-treelike proofs in our results.
Out of a total of 47,248%
 proofs,
29,338%
 of them are treelike.

\subsection{Results}

\Cref{tab:reordering-relative-stats} compares the relative improvement in space and
cumulative cost of each reordering policy compared to the original Metamath ordering,
averaged across \texttt{set.mm}. Despite \heurCumulPred{}'s relative simplicity,
\heurCumulPred{} and \heurCumulSU{} are very close in performance with respect to
cumulative cost. This suggests that simple heuristics for recursive branch heaviness are
effective for reducing cumulative cost.
For trees, (\heurNumSucc{}, \heurOriginal) defers to \heurOriginal{} since every
vertex has at most one successor. Overall, a mixture of \heurSU{} and either of
\heurCumulSU{} and \heurCumulPred{} seems to be the best approach for optimising
both space and cumulative cost.

\begin{table}[htbp]
  \caption{Average pebbling metrics for each reordering policy, broken down by treelike ($^\dag$) and non-treelike ($^\ast$) proofs.}
  \centering
  \begin{tabular*}{\textwidth}{@{\extracolsep{\fill}}lrrrrrrrr}
\toprule
Ordering policy & \multicolumn{2}{c}{$\Updelta_{\mathrm{sp}}^\dag$ (\%)} & \multicolumn{2}{c}{$\Updelta_{\mathrm{cc}}^\dag$ (\%)} & \multicolumn{2}{c}{$\Updelta_{\mathrm{sp}}^\ast$ (\%)} & \multicolumn{2}{c}{$\Updelta_{\mathrm{cc}}^\ast$ (\%)} \\
 & Median & Mean & Median & Mean & Median & Mean & Median & Mean \\
\midrule
Random & 0.00 & -1.03 & 0.00 & -0.62 & 0.00 & 0.04 & 2.96 & 0.74 \\
(SU, Original) & 0.00 & \textbf{12.48} & 0.00 & 11.10 & 22.86 & 23.12 & 21.99 & 23.12 \\
(CumulSU, Original) & 0.00 & 11.80 & \textbf{9.09} & 14.32 & 22.22 & 22.31 & 25.00 & 25.76 \\
(CumulPred, Original) & 0.00 & 11.53 & \textbf{9.09} & \textbf{14.36} & 22.22 & 22.09 & \textbf{25.14} & \textbf{25.81} \\
(NumSucc, Original) & 0.00 & 0.00 & 0.00 & 0.00 & 0.00 & 0.13 & 0.00 & 0.33 \\
(LastSucc, Original) & 0.00 & 10.16 & 1.64 & 11.42 & 14.29 & 14.23 & 14.29 & 15.61 \\
(SU, CumulSU, Original) & 0.00 & \textbf{12.48} & 8.33 & 14.05 & \textbf{24.14} & \textbf{23.58} & 24.14 & 24.99 \\
(SU, CumulPred, Original) & 0.00 & \textbf{12.48} & 8.33 & 14.05 & 24.00 & 23.56 & 24.14 & 24.98 \\
(SU, NumSucc, Original) & 0.00 & \textbf{12.48} & 0.00 & 11.10 & 22.86 & 23.13 & 22.02 & 23.16 \\
\bottomrule
\end{tabular*}

  \label{tab:reordering-relative-stats}
\end{table}

\Cref{tab:worse-performance-stats} breaks down the percentage of proofs
across \texttt{set.mm} where the heuristic policies perform worse than the original ordering. Recall that \heurSU{} is optimal for treelike proofs.
For non-treelike proofs, \Cref{fig:case-studies}~(d) presents a representative example showcasing an inherent limitation of DFS.
Note also that $\mathrm{AdjRate}$ is invariant for trees---this is because the adjacency rate of any (non-source) vertex's $n$ incoming edges is exactly $1/n$ regardless of the order in which DFS visits its predecessors.

\begin{table}[htbp]
  \caption{Percentage of proofs for which each policy performs worse than \heurOriginal{}, broken down by treelike ($^\dag$) and non-treelike ($^\ast$) proofs.}
  \centering
  \begin{tabular*}{\textwidth}{@{\extracolsep{\fill}}lrrrrrrrr}
\toprule
Ordering policy & \multicolumn{2}{c}{$\mathrm{sp}$ (\%)} & \multicolumn{2}{c}{$\mathrm{cc}$ (\%)} & \multicolumn{2}{c}{MID (\%)} & \multicolumn{2}{c}{AdjRate (\%)} \\
 & $^\dag$ & $^\ast$ & $^\dag$ & $^\ast$ & $^\dag$ & $^\ast$ & $^\dag$ & $^\ast$ \\
\midrule
Random & 20.85 & 34.33 & 29.87 & 42.84 & 29.87 & 41.18 & 0.00 & 22.20 \\
(SU, Original) & \textbf{0.00} & \textbf{2.04} & 0.62 & 4.14 & 0.62 & 5.25 & 0.00 & 18.41 \\
(CumulSU, Original) & 0.80 & 3.75 & 0.03 & 3.81 & 0.03 & 4.58 & 0.00 & 23.04 \\
(CumulPred, Original) & 1.06 & 3.80 & \textbf{0.00} & \textbf{3.66} & \textbf{0.00} & \textbf{4.41} & 0.00 & 23.37 \\
(NumSucc, Original) & \textbf{0.00} & 2.92 & \textbf{0.00} & 16.97 & \textbf{0.00} & 48.64 & 0.00 & \textbf{1.69} \\
(LastSucc, Original) & 1.28 & 8.97 & 2.22 & 11.88 & 2.22 & 11.92 & 0.00 & 15.83 \\
(SU, CumulSU, Original) & \textbf{0.00} & 2.52 & 0.46 & 4.31 & 0.46 & 5.28 & 0.00 & 21.76 \\
(SU, CumulPred, Original) & \textbf{0.00} & 2.49 & 0.46 & 4.27 & 0.46 & 5.26 & 0.00 & 21.74 \\
(SU, NumSucc, Original) & \textbf{0.00} & \textbf{2.04} & 0.62 & 4.26 & 0.62 & 8.21 & 0.00 & 18.37 \\
\bottomrule
\end{tabular*}

  \label{tab:worse-performance-stats}
\end{table}

\Cref{tab:lower-bound-attainment} reports the percentage of proofs where the theoretical
lower bound of pebbling space is attained.
Since \heurSU{} is optimal with respect to space,
it acts as a baseline for comparison for trees; the
wide margin of improvement over \heurOriginal{} (>30\%) highlights the
inherent inefficiency of the native Metamath format.
For non-treelike proofs, the best-performing policies achieve the lower bound almost six times as often as the original ordering.

\begin{table}[htbp]
  \caption{Percentage of proofs where space attains the theoretical lower bound.}
  \centering
  \begin{tabular}{lrr}
\toprule
Ordering policy & Treelike (\%) & Non-treelike (\%) \\
\midrule
Original & 54.57 & 3.74 \\
Random & 56.93 & 4.28 \\
(SU, Original) & \textbf{85.98} & 20.90 \\
(CumulSU, Original) & 82.75 & 15.85 \\
(CumulPred, Original) & 81.72 & 15.38 \\
(NumSucc, Original) & 54.57 & 3.80 \\
(LastSucc, Original) & 78.50 & 12.88 \\
(SU, CumulSU, Original) & \textbf{85.98} & 21.04 \\
(SU, CumulPred, Original) & \textbf{85.98} & \textbf{21.05} \\
(SU, NumSucc, Original) & \textbf{85.98} & 20.91 \\
\bottomrule
\end{tabular}

  \label{tab:lower-bound-attainment}
\end{table}

Finally, \Cref{tab:ordering-time-stats} ranks the ordering policies by their execution time.
Regardless of the policy, execution time across the entire \texttt{set.mm} library is less than four seconds. Even for very large proofs, step reordering is virtually instant.

\begin{table}[htbp]
  \caption{Execution times of each ordering policy across \texttt{set.mm}.}
  \centering
  \begin{tabular}{lrr}
\toprule
Ordering policy & Time per Statement (ns) & Time per Edge (ns) \\
\midrule
(SU, NumSucc, Original) & 2404.30 & 2139.58 \\
(SU, CumulSU, Original) & 2261.31 & 2012.33 \\
(SU, CumulPred, Original) & 1994.49 & 1774.89 \\
(LastSucc, Original) & 1913.14 & 1702.49 \\
(CumulSU, Original) & 1828.68 & 1627.33 \\
(SU, Original) & 1472.77 & 1310.61 \\
(CumulPred, Original) & 1294.81 & 1152.25 \\
(NumSucc, Original) & 1094.66 & 974.13 \\
Random & 801.13 & 712.92 \\
\bottomrule
\end{tabular}

  \label{tab:ordering-time-stats}
\end{table}

\section{Case Studies}
\label{sec:case-studies}

To show the effect of the heuristics, we begin with the previously mentioned
larger example \texttt{prmunb}, and then turn to smaller proofs whose
DAGs make the structural reason for improvements easy to see.
In the proof of \texttt{prmunb}, the
assumption \texttt{nnnn0} is introduced at the first step
of the proof even though it is first used only in step 48:
\begin{Verbatim}[samepage=true]
STEP   PREMISES   LABEL   EXPRESSION
1                 nnnn0   |- ( N e. NN -> N e. NN0 )
...
48     1,47       syl     |- ( N e. NN -> E. p e. Prime N < p )
\end{Verbatim}
Applying the ordering policy (\heurSU{}, \heurCumulSU{},
\heurOriginal{}) delays \texttt{nnnn0} until step 47, reducing
the space from 9 to 6 and
cumulative cost from 224 to 137. In the rest of this section we present
smaller examples where improvements are visually apparent.

\begin{figure}[htbp]
    \centering
    \makebox[\textwidth][c]{
      \begin{minipage}[t]{0.15\textwidth}
        \centering
        \includegraphics[width=\linewidth]{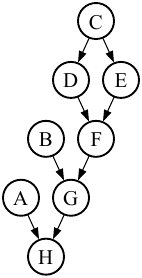}
        {\small\textbf{(a)}}
      \end{minipage}
      \hspace{\fill}
      \begin{minipage}[t]{0.20\textwidth}
        \centering
        \includegraphics[width=\linewidth]{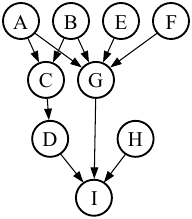}
        {\small\textbf{(b)}}
      \end{minipage}
      \hspace{\fill}
      \begin{minipage}[t]{0.24\textwidth}
        \centering
        \includegraphics[width=\linewidth]{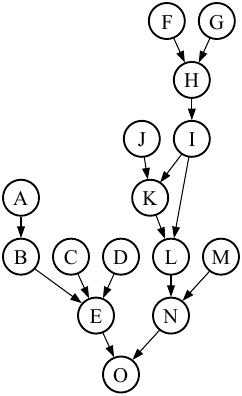}
        {\small\textbf{(c)}}
      \end{minipage}
      \hspace{\fill}
      \begin{minipage}[t]{0.23\textwidth}
        \centering
        \includegraphics[width=\linewidth]{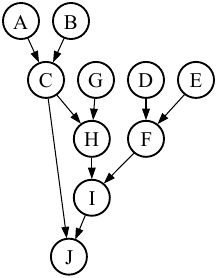}
        {\small\textbf{(d)}}
      \end{minipage}
      }
    \caption{Proof DAGs for theorems \texttt{eldisjs5}, \texttt{subcan}, \texttt{neifval}, and \texttt{knoppndvlem3}, with vertices relabelled alphabetically according to the original Metamath step order.}
    \label{fig:case-studies}
\end{figure}

Consider \Cref{fig:case-studies}~(a). Inference $A$ is executed 7
steps before it is required by inference $H$. Inference $B$ is also
executed unnecessarily early. The policy (\heurSU{}, \heurCumulSU{},
\heurOriginal{}) instead computes the ordering
$(C, D, E, F, B, G, A, H)$, executing inferences $B$ and $A$ just as they are
required, i.e., the edges $(A, H)$ and $(B, G)$ now have unit index
distance. Pebbling space is reduced from 4 to the theoretical lower
bound of 2, and cumulative cost is reduced from 20 to 12.

Adjacency rate and pebbling space are not always positively
correlated, however. In \Cref{fig:case-studies}~(b), the policy
(\heurSU{}, \heurCumulSU{}, \heurOriginal{}) computes the ordering
$(A, B, E, F, G, C, D, H, I)$. The original ordering has four edges
with unit index distance: $(B, C), (C, D), (F, G)$, and $(H, I)$. The
new ordering, however, has only three: $(F, G), (C, D)$, and $(H,
I)$. Overall, the MID worsens from 2.6 to 2.7. Despite this, space
improves from 5 to 4 and cumulative cost from 24 to 21. Thus there are
cases where tradeoffs between memory consumption and index distance
may occur.

Recall from \Cref{tab:space-vs-cost} that cumulative cost can improve
even when space is unchanged. \Cref{fig:case-studies} (c) shows a more
dramatic example where space actually worsens. The policy (\heurSU{},
\heurCumulSU{}, \heurOriginal{}) computes
    $$
      (\underbrace{F, G, H, I, J, K, L, M, N}_{\text{right branch}},
        \underbrace{A, B, C, D, E}_{\text{left branch}}, O) .
    $$
This is identical to the original ordering except that the right
branch ending in $N$ is traversed before the left branch ending in
$E$.
In this case, cumulative cost improves from 31 to 27 even though space
worsens from 3 to 4.
This is possible because in the original ordering,
$\varepsilon(E)$ must be held in memory while the larger branch is traversed.
In the new ordering,
$\varepsilon(N)$ is held in memory for far fewer steps,
although it does force the bottleneck configuration $\{ N, B, C, D\}$.
Arguably, a brief spike in configuration cost could be a small price
to pay for a significantly reduced cumulative cost.

Finally, \Cref{fig:case-studies} (d) presents a rare
example (4.31\%) where the
policy (\heurSU{}, \heurCumulSU{}, \heurOriginal{}) computes a
worse ordering $(A, B, C, G, H, D, E, F, I, J)$. Space is
inflated from 3 to 4, and cumulative cost from 20 to 21. This is a
limitation of DFS: predecessors of different vertices cannot be
interleaved out of order. Thus the edge $(G, H)$ must be traversed
before any predecessors of $F$ can be visited. In this case, the
bottleneck configuration for space is $\{ C, G, D, E \}$.

\section{Future Work}

Future directions include adapting proof DAGs to account for backwards reasoning,
which could possibly be modelled using \emph{black-white} pebbling games.
This may enable adapting our methods for application in other mathematical
libraries such as Mizar~\cite{mizar_2010}, Lean's Mathlib~\cite{lean_mathlib_2020},
and Isabelle/HOL's AFP~\cite{afp_repository}.
To capture variations in the complexity of mathematical expressions,
future work could explore using \emph{weighted} pebbling games and
assign higher weights to expressions with more symbols
or deeper nested structure.

\section{Conclusion}

Using pebbling games, we studied how proof step reordering in Metamath influences the memory consumption of proof verification. Our heuristic DFS algorithms improve peak and cumulative memory consumption on many proofs in Metamath's ZFC set theory library (\texttt{set.mm}) while preserving proof length. Case studies illustrate how these improvements arise and where they may fail, e.g., as an inherent limitation of DFS. Overall, our results support proof reordering as a practical optimisation target for legibility proxy metrics, and motivate future evaluation on other formal libraries beyond Metamath.

\begin{credits}
\subsubsection*{\ackname}
This research was supported by an Australian Government Research Training Program Scholarship and the Renaissance Philanthropy grant Deeper.

\subsubsection*{\discintname}
The authors have no competing interests to declare that are relevant to the content of this article.
\end{credits}

\bibliographystyle{splncs04}
\bibliography{references}

\end{document}